# A CUSTOMIZED FLOCKING ALGORITHM FOR SWARMS OF SENSORS TRACKING A SWARM OF TARGETS.

Anupam Shukla, Gaurav Ojha, Sachin Acharya, Shubham Jain[1]

[1]Department of Information and Communication Technology, Indian Institute of Information Technology and Management, Gwalior – 474010, INDIA

## ABSTRACT

*Wireless mobile sensor networks (WMSNs) are groups of mobile sensing agents with multi-modal sensing capabilities that communicate over wireless networks. WMSNs have more flexibility in terms of deployment and exploration abilities over static sensor networks. Sensor networks have a wide range of applications in security and surveillance systems, environmental monitoring, data gathering for network-centric healthcare systems, monitoring seismic activities and atmospheric events, tracking traffic congestion and air pollution levels, localization of autonomous vehicles in intelligent transportation systems, and detecting failures of sensing, storage, and switching components of smart grids.*

*The above applications require target tracking for processes and events of interest occurring in an environment. Various methods and approaches have been proposed in order to track one or more targets in a pre-defined area. Usually, this turns out to be a complicated job involving higher order mathematics coupled with artificial intelligence due to the dynamic nature of the targets. To optimize the resources we need to have an approach that works in a more straightforward manner while resulting in fairly satisfactory data. In this paper we have discussed the various cases that might arise while flocking a group of sensors to track targets in a given environment. The approach has been developed from scratch although some basic assumptions have been made keeping in mind some previous theories. This paper outlines a customized approach for feasibly tracking swarms of targets in a specific area so as to minimize the resources and optimize tracking efficiency.*

## 1. INTRODUCTION

Flocking behavior is the behavior exhibited when a group of birds, called a flock, are hunting or wandering in a flight. There are parallels with the shoaling behavior of fish, the swarming behavior of insects, and herd behavior of land animals and so on. Computer simulations and mathematical models which have been developed to emulate the flocking behaviors of birds can generally be applied also to the "flocking" behavior of other species [1]. As a result, the term "flocking" is sometimes applied, in computer science, to species other than birds. In general terms, flocking algorithms are applied to objects – a group of which is then referred to as swarms. Objects may include satellites, sensors, imaging devices, mobile robots, or simply any singular item which is required to communicate with its replicas and form groups.
In flocking simulations, there is no central control; each object behaves autonomously. In other words, each object has to decide for itself which flocks to consider as its environment. Usually environment is defined as a circle (2D) or sphere (3D) with a certain radius (representing reach). A basic implementation of a flocking algorithm has complexity $O(n^2)$ - each object searches through all other objects to find those which fall into its environment.

The main problem of interest is distributed multi-target tracking using multi mobile wireless sensor networks. Sensor networks can be divided into two categories based on the relative sensing range of their nodes: 1) Limited Sensing Range (LSR) sensor networks and 2) Unlimited Sensing Range (USR) sensor networks. Fig. 1 illustrates the difference between wireless sensor networks with limited sensing range and unlimited sensing range. Sensors with unlimited (or very long) sensing range include radars and GPS and sensors with limited sensing range include LIDAR, cameras, IR sensors, and sonars.

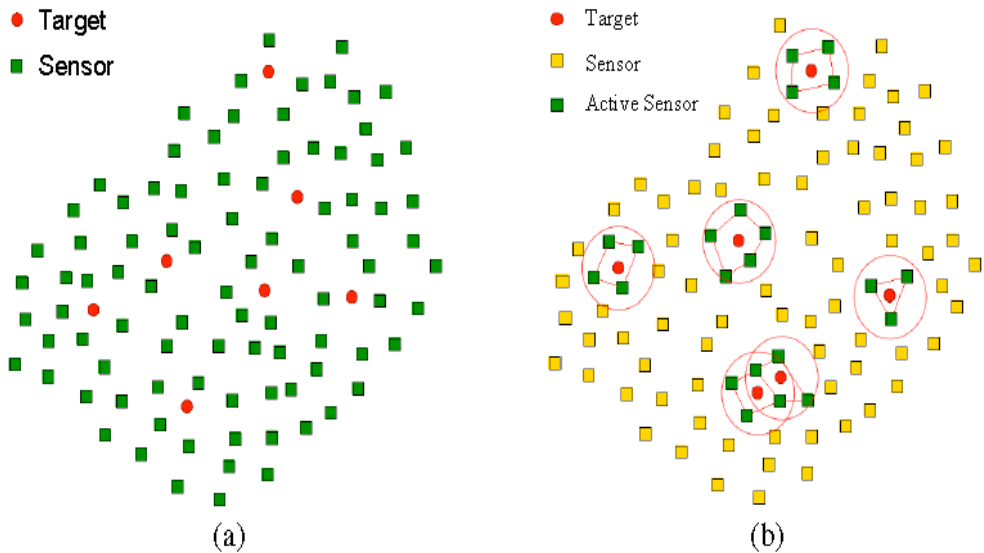

Figure 1. USR and LSR sensor networks: (a) Unlimited sensing range sensor network and (b) Limited sensing range sensor network. In (a), all sensors are active and observe all targets manoeuvring in the sensing field. In (b), each target is only observed by a limited set of active sensors that are within the sensing range around the target.

In an unlimited sensing range sensor network all sensors have the capability of observing a target manoeuvring in the sensing field for all t > 0. However, most real-life sensor networks are the LSR-type due to their bounded sensing range and cheap. The main challenge in LSR sensor networks is that targets of interest cannot be observed by all the sensors and therefore high performance estimation and tracking of the targets requires reaching a consensus on state of the targets among the collaborative mobile sensors. If sensors make low quality measurements, they will not be able to accurately track targets. To obtain quality estimates, the sensors should self-locate themselves closer to the targets and to get into close proximity of targets. Staying close to targets requires knowing target location with high levels of certainty. For accurate target location, dynamic placement of sensors near targets is required. In other words, dynamic placement of sensors and high quality estimation and tracking of targets are coupled problems this is schematically shown in figure 1. From figure1 it is clear that to obtain quality estimates, the sensors should be placed closed proximity of the target. In Figure 2 illustrates the importance of sensor placement for sensor networks with limited sensing range. As sensors approach targets, the information value of measurements increases. The scenario in Figure 2 (c) represents the case in which sensors are placed in the sensing field such that all targets are observed by at least one sensor.

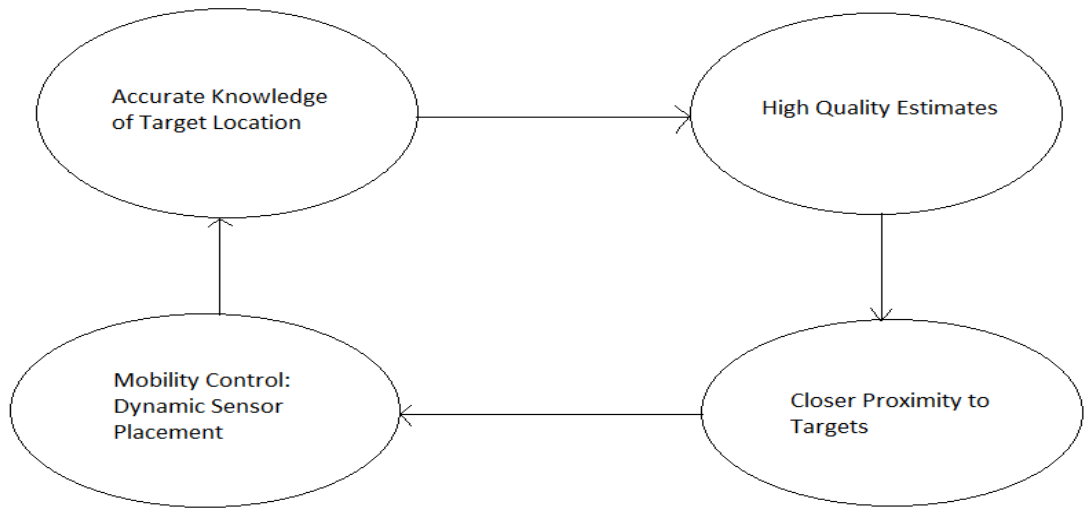

Figure 2. Dynamic sensor placement, estimations, tracking of the target should self-locate themselves closer to the targets and to get into close proximity of targets sensors need to have a proper dynamic placement, i.e. control-driven estimation. This mobility control needs accurate knowledge of target location. To better control the mobility of the sensors, the sensors need to accurately estimate the location of the targets.

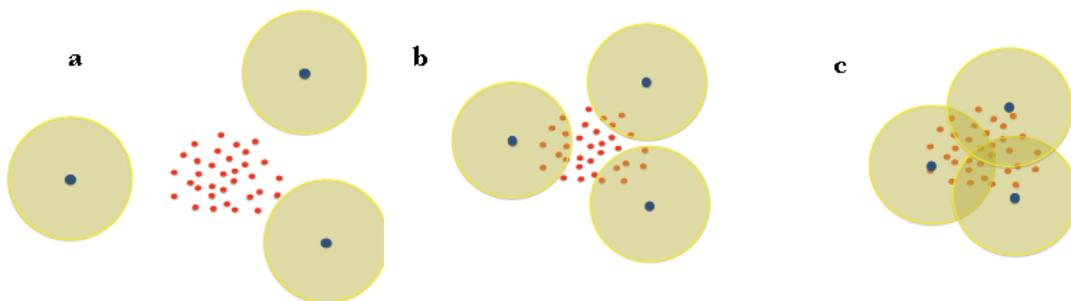

Figure 3. Three sensor placements scenarios: (a) none of the targets are observed by sensors and the, (b) a large portion of targets are missed due to lack of proper placement of the sensors, and (c) sensors are observing all targets and have a better placement compared to both cases (a) and (b). in this figure blue dot denotes the mobile robots carrying sensor(Limited Sensing Range (LSR) sensor), yellow bordered circle is the sensing range, and red dots are the targets.

## 2. LITERATURE REVIEW

Traditionally, a lot of highly advanced mathematical techniques have been used to design flocking algorithms and understand the behavior of swarms. But the situation becomes more complex when we need to track certain targets which may also exhibit characteristics of flocking. Some basic methods that have been used over the years are 1. Kalman filtering and its variations, 2. Alpha-beta filtering, 3. Kernel adaptive filtering and so on.

Kalman filtering algorithms and theory have been one of the most effective ways to address estimation and multi-target tracking problems. The algorithm works in a two-step process. In the prediction step, the Kalman filter produces estimates of the current state variables, along with their uncertainties. Once the outcome of the next measurement (necessarily corrupted with some amount of error, including random noise) is observed, these estimates are updated using a weighted average, with more weight being given to estimates with higher certainty. Because of the algorithm's recursive nature, it can run in real time using only the present input measurements and the previously calculated state; no additional past information is required.

Variations of Kalman filtering algorithm include a. The distributed Kalman filtering technique (DKF) [8], b. Kalman Consensus Filtering (KCF), c. Hybrid Kalman filtering and d. Kalman-bucy filtering.

An Alpha-beta filter is a simplified form of observer for estimation, data smoothing and control applications. It is closely related to Kalman filters and to linear state observers used in control theory. Its principal advantage is that it does not require a detailed system model.

Kernel adaptive filtering is an adaptive filtering technique for general nonlinear problems. It is a natural generalization of linear adaptive filtering in reproducing kernel Hilbert spaces. Kernel adaptive filters are online kernel methods, closely related to some artificial neural networks such as radial basis function networks and regularization networks. Some distinguishing features include: The learning process is online, the learning process is convex with no local minima, and the learning process requires moderate complexity.

Quite a few researches have been done in this field and researchers have come up with a variety of techniques. Reynolds [12] has described the motion of flocks as a result of three basic principles viz. 1. Cohesion, which is an intrinsic attempt to stay close to other flock mates, 2. Alignment, where the flock mates attempt to match their velocities and 3. Separation, meaning that collision has to be avoided amongst themselves. Vicsek [11] gave a flocking model in which the headings of the mobile agents are controlled in such a way that they match the average of the headings of their closest neighbours. Spears [10] used artificial physics to provide a distributed control of large number of agents and hence achieved certain positive results such as self-assembly, self-repair and fault-tolerance. In more recent researches, it has also been considered that the agents may flock in given space with a virtual leader [9] to guide them. Though this approach did not consider a target system in the flocking problem.

Area exploration is also an important part of the problem and a novel approach towards the same has been discussed in [2] and [3] addresses some issues of significance too.

In [4], Christoph, Thomas, et. al have discussed the minimalist flocking algorithm for swarm robots. By this algorithm, simple swarm robots works without communication, memory or global information and analyses its potential of aggregating an initially scattered robot swarm. But the mobility of swarm robots depends on flock's size, increasing the flock size decreases the mobility.

In [5], Weishi, Hong, et. al have discussed a group tracking of flock targets in low-altitude airspace. In this approach, encouraging results are obtained with high-efficiency and the data association problem is also overcome. But this method doesn't take into consideration the influence each object has on the others during flight and so it becomes difficult to adjust its velocity.

In [7], Yan, Naixue, Nak Young, et. al have discussed a decentralized and adaptive flocking algorithm for autonomous mobile robots. This algorithm can avoid the collision among robots, its neighbors and obstacles in the environment. This algorithm would be very useful in practical applications, like rescue after earthquake and space exploration.

The biggest challenge to solving this problem is to decide how to align the sensors so as to optimize the resource usability while at the same time track targets efficiently without missing on any of them. Though this approach involves complex mathematics, we present it in an easily assimilable form so that it provides a stable foundation for further research to be performed in this area.

## 3. PROPOSED METHODOLOGY

In the proposed methodology, we begin with some assumptions regarding the positions and movements of the sensor robots and the targets.

To start with, we assume that the area to be tracked by the sensor robots is fixed. That means, the dimension and span of the area are known well-in advance. Vehicles or targets to be tracked enter from one specific direction and move towards the opposite direction but are to be tracked as long as they are in the specified area. Each sensor has a fixed range which extends in the form of circular area around the sensor. This can be termed as its 'zone'. A sensor can only track objects within its 'zone'. For our specific customized algorithm, each sensor possesses two kinds of zones. A primary zone and a secondary zone. Primary zone is defined as the zone created by that part of the sensor which can track targets, i.e. a high frequency signal is present in this zone which can track the targets efficiently. A secondary zone is then defined as an extended area outside the primary zone wherein low frequency signals, sufficient to track other sensors, is present.

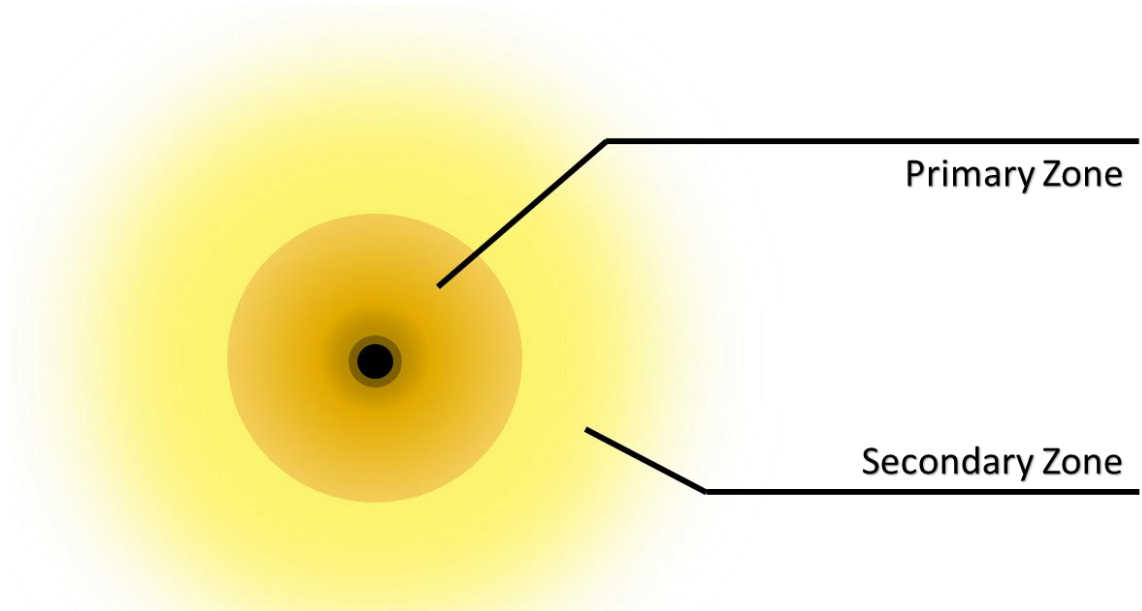

Figure 4. Primary and secondary zones in our hybrid sensors.

Activation of the secondary zone is subject to requirements and needs. The primary zone, too, can sense the location of nearby sensors but in our cases it is not required. Keeping the secondary zone deactivated optimizes the resource requirements to a large extent.

Next, we define the following customized approach for tracking a swarm of targets –

We first define an area 'A' which represents the tracking area, i.e. the area within which we are interested in tracking that swarm of objects.

The radius of the range of each sensor (S) is 'r', and n(S) represents the number of sensors that are there for any given instance.

We start by finding out the total area covered by 'n' sensors and denote it by $A_S$.

So, $A = l \times b$ where $l$ is the length of the ground and $b$ is its breadth.

Range or zone of each sensor is defined as $A_s = \pi r^2$

**Case 1:** $A_S \geq A$ so n(S) is sufficient to cover the area A. Sensors are arranged so as to cover the whole area and then tracking is done using suitable handoff techniques.

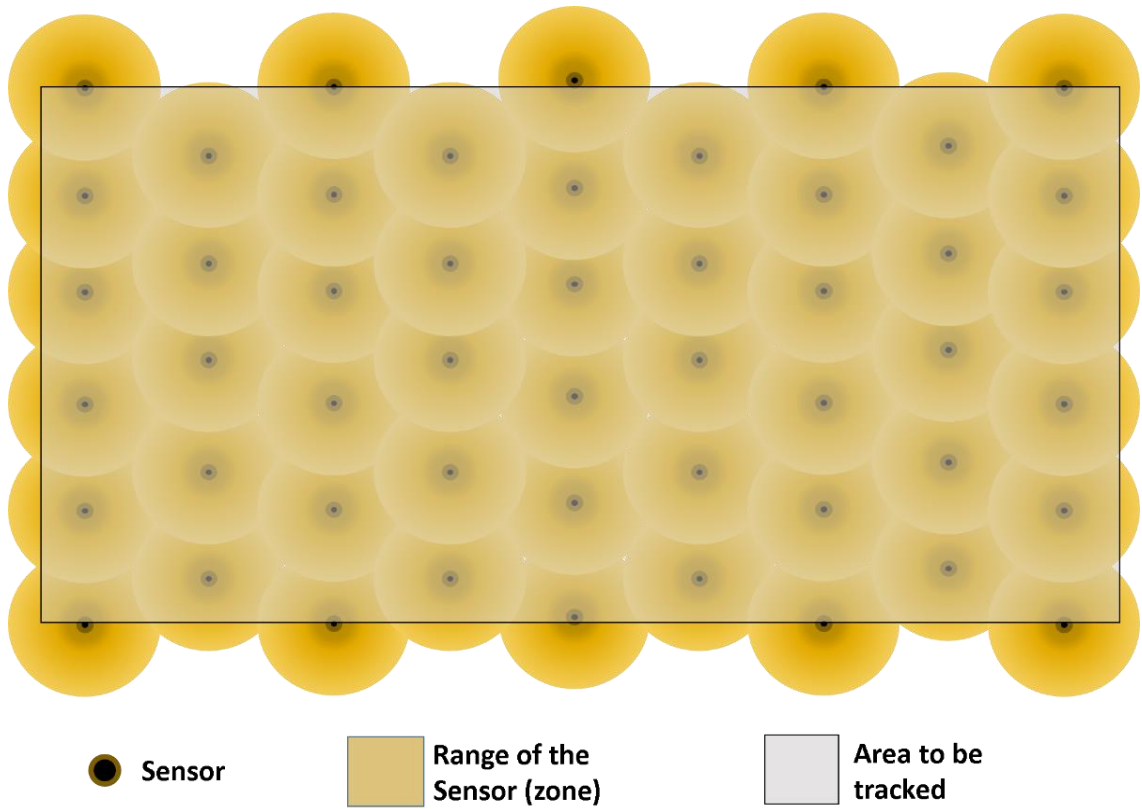

Figure 5. Case 1, when the number of sensors is enough to cover the region to be tracked.

In this case, there are n sensors covering the given area. Since the number of sensors are n, the total area covered by the sensor is given by $n\pi r^2$.

Therefore, if $n\pi r^2 \geq l \times b$ then, a suitable handoff technique is used to track the moving targets.

As the targets move, a hand-off technique can be applied. If the Velocity of the target in the outward direction is either increasing or constant and the distance of the target from the sensor is increasing, then a hand-off technique can be used to pass off the information of the location of the target to the nearby sensor. A customized handoff algorithm can be further developed for this approach based upon the study outlined in [13] where it has been discussed about an approach where handoff is based on relative zonal strengths between sensors and targets and it occurs when the relative zonal strength between a sensor and the target decreases while at the same time it increases with respect to another sensor.

As shown in figure, the sensors have arranged themselves to cover the whole area. Some sensors may be left out after this exercise because we have assumed that their number is sufficient to cover the whole area.

**Case 2:** $A_s \gg A$, so we can put some of the sensors to rest and then do Case 1.

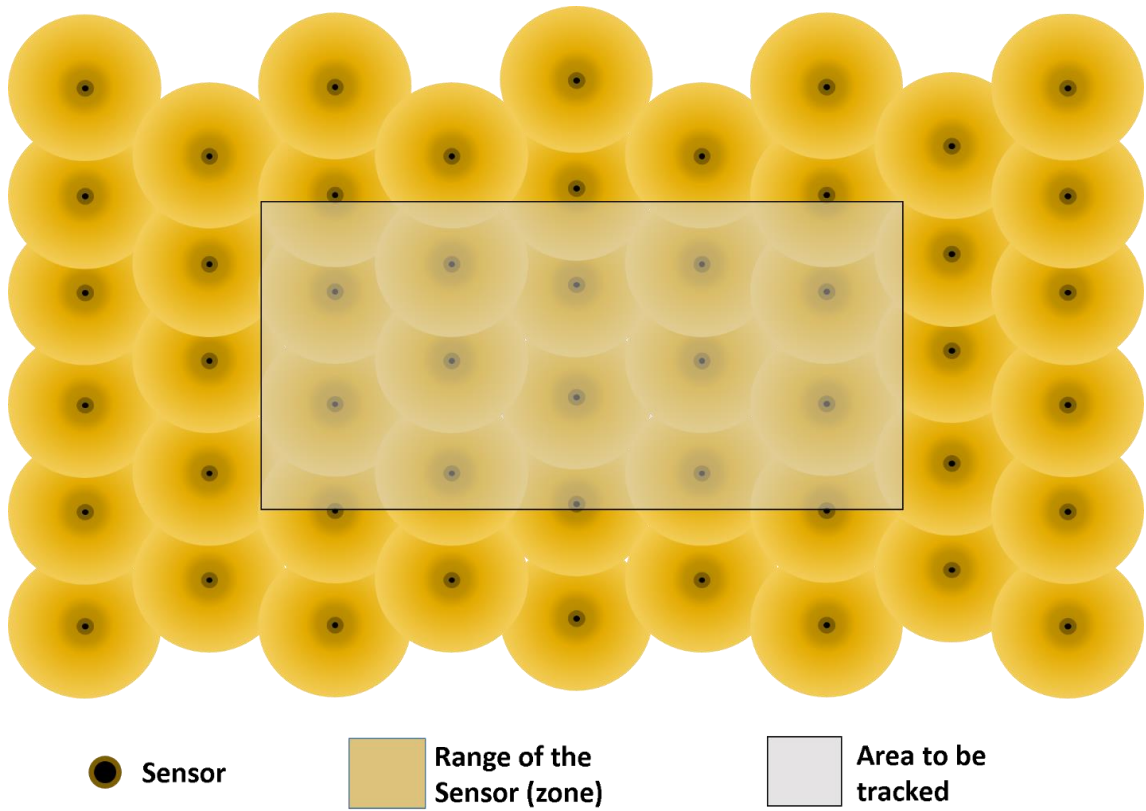

Figure 6. Case 2, when the number of sensors is more than enough to cover the region to be tracked.

In this case, since the number of sensors is extremely large, the area covered by the sensors is extremely greater than the area of the ground. In this situation, the excess sensors may be put to some other use and Case 1 may be applied.

Case 3: AS < A, we arrange the sensors in the area towards the direction from where the vehicles are supposed to enter and then use a suitable algorithm to define the motion of the sensors with respect to the targets.

This is a tricky case. We take into consideration the approach defined by Xiaoyuan, et al. [6] and for the dynamic system, the following assumptions have to be made –

1. Each sensor knows its own absolute position in terms of a co-ordinate system defined for the area in which tracking is to be done.
2. Each sensor has information about the positions of only those sensors which are located in its secondary zone. If the secondary zones of two sensors overlap, they still know each other's positions and velocities because each knows its own absolute position.
3. Each sensor has the information about the positions and velocities of the targets in its primary zone only.
4. Each sensor can track only a limited number of targets.

As previously defined, the area to be tracked is A and the number of sensors is n(S), where As is the total area that can be covered by the primary zones of the sensors and As < A. Now, let's say that n(s) is sufficient to cover $\frac{A}{k}$th part of the total area A using the approach defined in case 1. So, we arrange n(S)/2 sensors in A/2kth part of the area. The remaining n(S)/2 sensors then evenly distribute themselves in the remaining $\frac{A(2k-1)}{2k}$ part of the area such that their secondary zones overlap and they know the positions of each other pairwise. If their secondary zones cannot overlap, then it boils down to case 4 implying that the number of sensors is just not sufficient to cover the whole area and the targets cannot be tracked.

Considering that the case holds true, we can now define the following behaviour for the sensors-
1. As long as the targets move in the first A/2k part of the area, they can be efficiently tracked without any sensor movement as explained in case 1.
2. When a target enters the next $\frac{A(2k-1)}{2k}$ part of the area, the sensors use a 'pass the parcel' approach and keep handing off the tracking job to the nearest sensors towards which the targets are moving based upon their velocities and direction of movement.

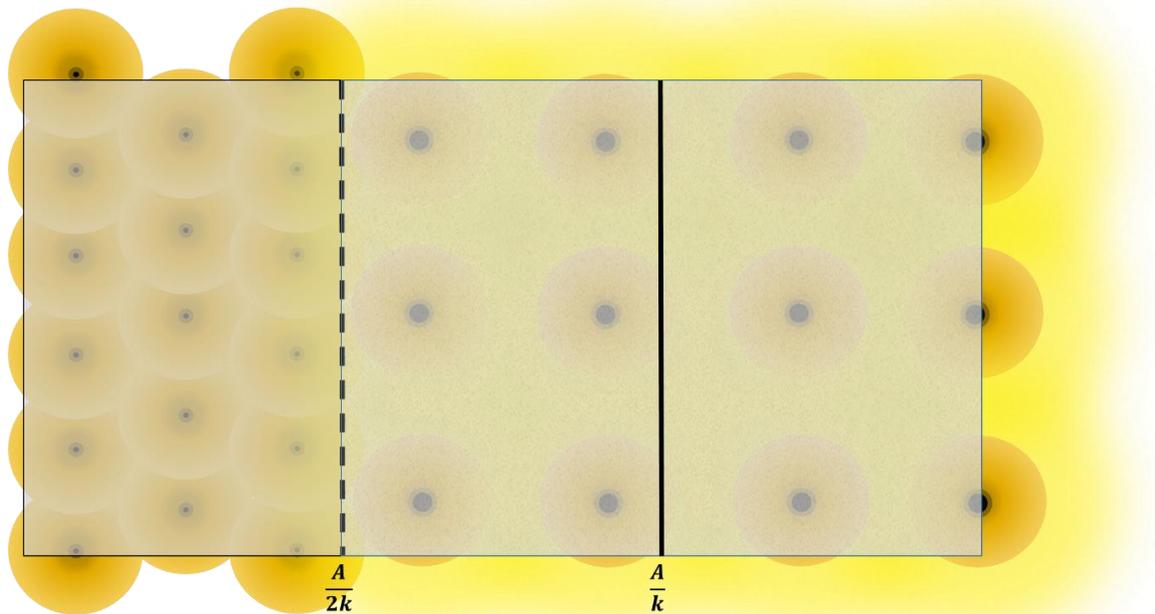

☐ $\frac{A}{2k}$ Part of the area (covered by primary zones)  ☐ $\frac{A(2k-1)}{2k}$ Area covered in secondary zones

Figure 7. Case 3, when the number of sensors is just optimum to cover the region to be tracked.

**Case 4:** AS << A, the number of sensors is insufficient to cover the area given.
In this case, the number of sensors is insufficient to cover the area. Therefore, if $n\pi r^2 \leq l \times b$, then we cannot effectively track the targets in the given area and thus tracking will not be possible.

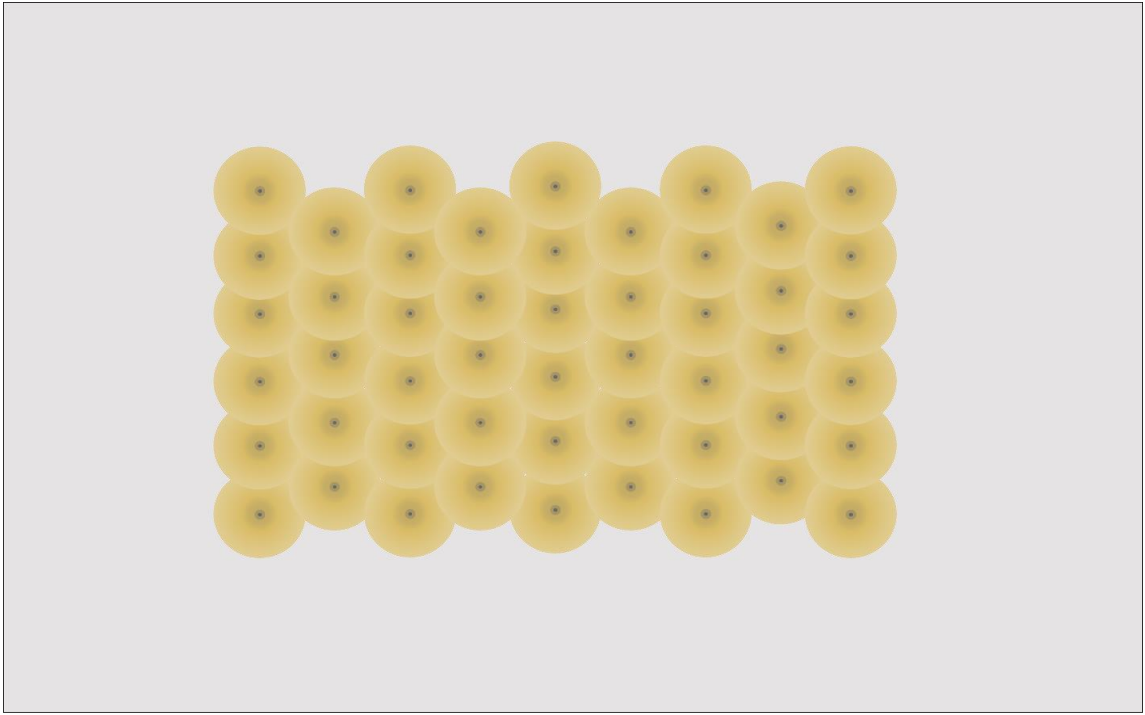

| | | Range of the | | Area to be |
| - | - | - | - | - |
| | Sensor | Sensor (zone) | | tracked |

Figure 8. Case 4, when the number of sensors is too less than what is required to cover the region in which tracking is to be done.

## 4. EVALUATION AND INFERENCES

We built a customized application to test the feasibility and working of the proposed algorithm. We have obtained some brilliant results. We have predicted the optimum number of sensors required for a given area and assigning it to one of the cases as described above. The results of the evaluation are given as follows –

| Area | Radius of Primary Zone | Radius of Secondary Zone | Area of Primary Zone | Area of Secondary Zone | No. of Sensors for Optimized Tracking | Value of K | Case Used - |
| --- | --- | --- | --- | --- | --- | --- | --- |
| 4 | 2 | 4 | 12.566 | 50.265 | 1 | 0.32 | Case 1 |
| 10 | 2 | 4 | 12.566 | 50.265 | 1 | 0.8 | Case 3 |
| 20 | 2 | 4 | 12.566 | 50.265 | 2 | 0.8 | Case 3 |
| 40 | 2 | 4 | 12.566 | 50.265 | 4 | 0.8 | Case 3 |
| 50 | 2 | 4 | 12.566 | 50.265 | 4 | 0.99 | Case 3 |
| 80 | 2 | 4 | 12.566 | 50.265 | 6 | 1.06 | Case 3 |
| 90 | 2 | 4 | 12.566 | 50.265 | 6 | 1.19 | Case 3 |
| 160 | 2 | 4 | 12.566 | 50.265 | 10 | 1.27 | Case 3 |
| 210 | 2 | 4 | 12.566 | 50.265 | 12 | 1.39 | Case 3 |
| 250 | 2 | 4 | 12.566 | 50.265 | 14 | 1.42 | Case 3 |
| 1000 | 1 | 3 | 3.142 | 28.274 | 80 | 3.98 | Case 3 |

| 1000 | 2 | 4 | 12.566  | 50.265  | 54 | 1.47 | Case 3 |
| 1000 | 3 | 5 | 28.274  | 78.54   | 40 | 0.88 | Case 3 |
| 1000 | 4 | 6 | 50.265  | 113.097 | 32 | 0.62 | Case 3 |
| 1000 | 5 | 7 | 78.54   | 153.938 | 28 | 0.45 | Case 1 |
| 1000 | 6 | 8 | 113.097 | 201.062 | 24 | 0.37 | Case 1 |
| 1000 | 7 | 9 | 153.938 | 254.469 | 20 | 0.32 | Case 1 |

The value of K was calculated keeping in mind that the optimum number of sensors required (as described by Case 3 of the proposed methodology) are found out by equating the area covered by primary zones of half of the sensors to the area covered by the secondary zones of the other half of the sensors.

So, we got $K = \frac{A}{n\pi r^2}$ where A represents the area of the region to be tracked and n is the number of sensors and r is the radius of the primary zones.

The Value of n is calculated using the formula $n = \frac{2A}{\pi(R^2 - r^2)}$ where R is the radius of the secondary zone.

Based on an extended version of the above table, we plotted the following graphs –

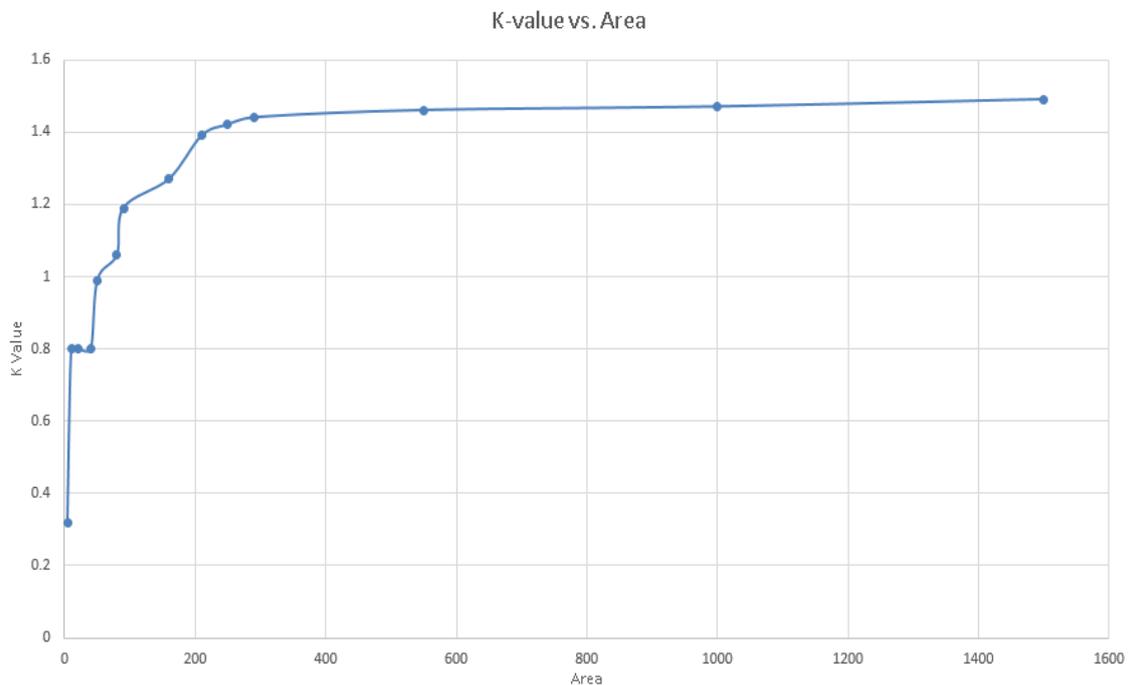

Figure 9. Comparison of K-values v/s Area of the region to be tracked.

The above graph shows the variation of k-value as the area increases given that the radii of the primary and secondary zones are constant. The results show that the value of K converges to a constant even though area increases steadily. We can say that the constant K value indicates a general tendency for the system to use Case 3.

Next, we have plotted a graph of K-value v/s Radius of the Primary Zone.

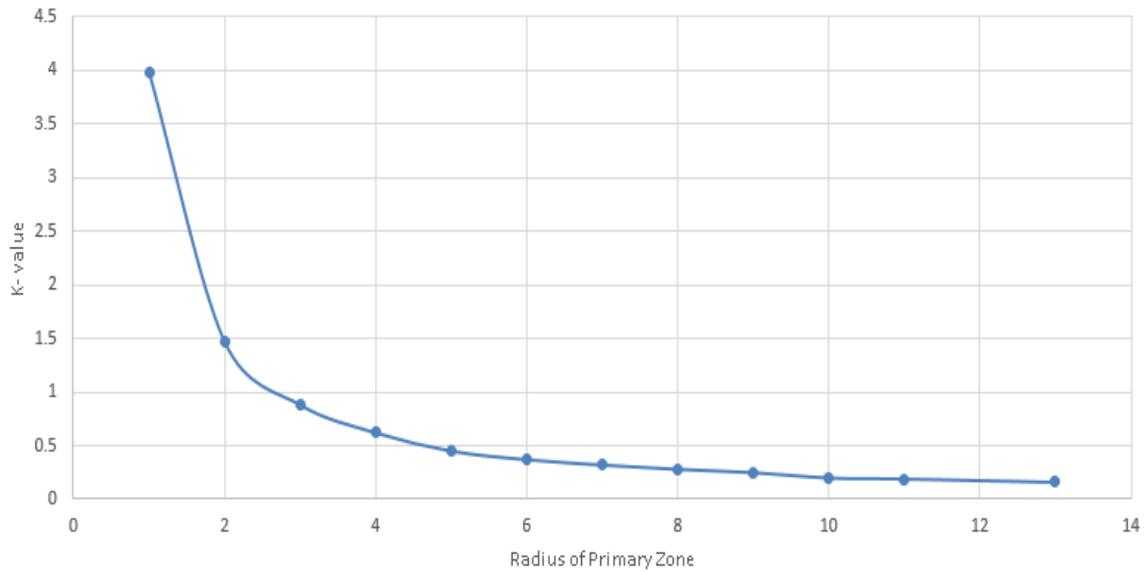

Figure 10. Comparisons of K-values v/s Radius of primary zone of the sensor.

We found out that, keeping the area constant, the K-value diminishes continuously with increase in the radius of the primary zone. Beyond k = 0.5, the system as enough sensors to cover the whole area and so, it utilizes the approach mentioned in case 1.

Finally, we plotted a graph of number of sensors required v/s area of the region to be tracked. It was observed that the number of sensors varies directly as the area.

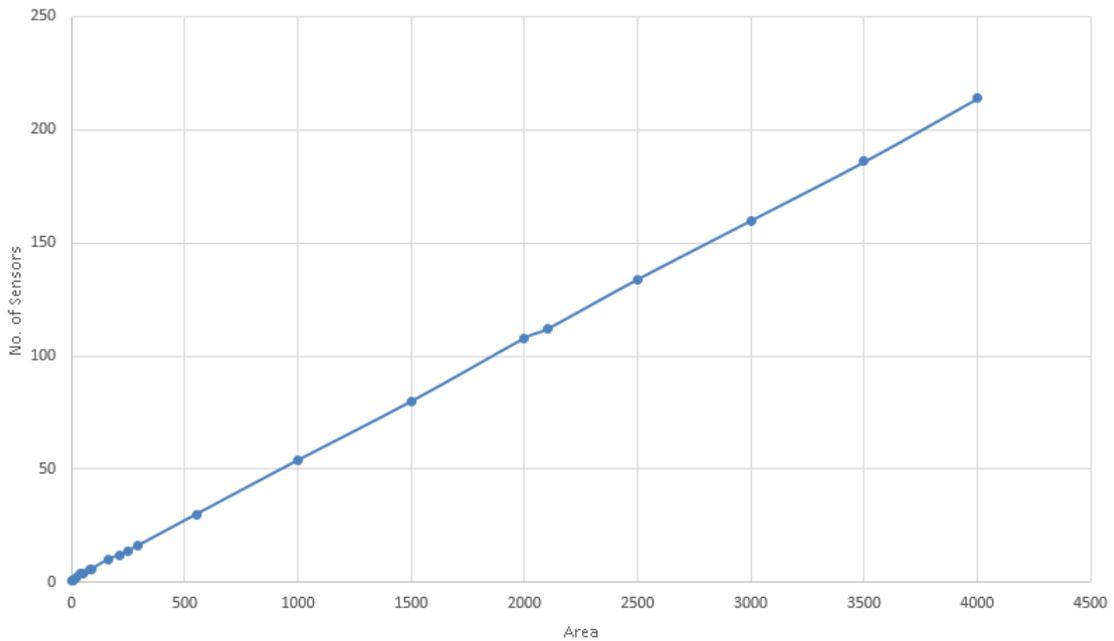

Figure 11. Comparison of Area v/s Number of sensors required to track the area.

## 5. APPLICATIONS

Sensor networks have a wide range of applications in security and surveillance systems, environmental monitoring, data gathering for network-centric healthcare systems, monitoring seismic activities and atmospheric events, tracking traffic congestion and air pollution levels, localization of autonomous vehicles in intelligent transportation systems, and detecting failures of sensing, storage, and switching components of smart grids.

In security and surveillance systems, it can be used to track infiltrations and movements of high profile targets or VIPs for their security. It can be a very efficient method to provide security at the border of countries. In environmental monitoring, it can used effectively to track the motion of endangered animals to make sure they're in a safe zone. It can also be used to see if any harmful wild animals or poachers are approaching any restricted areas viz. human habitations and protected areas respectively.

In traffic congestion monitoring this method can be very accurate at pulling out details about jam-prone areas and help chalk out techniques to prevent congestion in real time.

## 6. CONCLUSION

We have developed a customized approach for tracking a swarm of targets using a swarm of sensors. We have also taken a number of cases which can possibly arise in the implementation of this technique. It can prove to be very effective in a number of cases as discussed in the applications.

But we have also seen that the method fails if the area to be tracked is really large or the number of sensors is small. In that case one approach is to let the sensors move haphazardly behind the targets and keep tracking them as long as possible. In this approach we face the problem of de-flocking of the targets. We can never be very sure that a flock of targets won't separate while moving. Another approach is to assign the sensors some specific targets so that they don't get overloaded. But this approach requires previous information about the movement of the targets which may not always be possible.

Further research is required in order to develop more advanced techniques for case 3 described in the proposed methodology and we also need to address the issue of failure of the technique as outlined in case 4. A better hand-off technique tailor-made to suit this approach can be designed to take full advantage of this customized flocking algorithm.

## REFERENCES

[1] Basics of Flocking Behaviour http://ccl.northwestern.edu/netlogo/models/Flocking and simulations http://www.red3d.com/cwr/boids/

[2] Shukla, Anupam, et al., A novel approach for path planning, area explore and area retrieval of an autonomous robot using the low-cost infrared sensors, 7[th] IEEE Conference on Industrial Electronics and Applications (ICIEA), 2012.

[3] Tiwari, Ritu, Sharma, Sanjeev, et al., Area Exploration by Flocking of Multi Robot, Procedia Engineering, Volume 41, 377-382, published by Elsevier, 2012.

[4] Christoph, M., Thomas, S. and Karl, C., A minimalist flocking algorithm for swarm robots, Advances in Artifial Life, Darwin meets von Neumann, published in Springer LNCS, Volume 5778, pp 375-382, 2011.

[5]    Weishi, C., Hong, L., et al., 'Group Tracking of Flock Targets in Low-Altitude Airspace', in Proceedings of the 9th IEEE International Symposium on Parallel and Distributed Processing with Applications Workshops, 131-136, May 26-28, 2011.

[6]    Xiaoyuan, L., et al., Flocking algorithm with multi-target tracking for multi-agent systems, Pattern Recognition Letters, Volume 31, 800-805, published by Elsevier, 2010.

[7]    Yan, Y., Naixue, X., et al., A decentralized and adaptive flocking algorithm for autonomous mobile robots, 3rd International Conference on Grid and Pervasive Computing – Workshops, March 2008

[8]    R. Olfati-Saber, Distributed Kalman Filtering for Sensor Networks, In Proceedings of the 46th IEEE Coneference on Decision and Control, New Orleans, LA, USA, Dec. 12-14, 2007.

[9]    Shi, H. et al., 2007, Flocking of multi-agent systems with a virtual leader, In Proceedings of the 2007 IEEE Symposium on Artificial Life, Hawaii, USA, pp. 287–294.

[10]    Spears, W. et al., 2004, Distributed, physics-based control of swarms of vehicles, Autonomous Robots 17, 137–162.

[11]    Vicsek, T. et al., 1995, Novel type of phase transition in a system of self-driven particles. Phys. Rev. Lett. 75 (8), 1226–1229.

[12]    Reynolds, C. W., 1987, Flocks, birds, and schools: A distributed behavioral mode. Comput. Graphics (ACM SIGGRAPH '87 Conf. Proc.) 21 (6), 25–34.

[13]    Nasif, Ekiz, Salih, Tara, et. al., 'An Overview of Handoff Techniques in Cellular Networks', International Journal of Information Technology, Volume 2, Number 2.

**Authors**

Prof. Anupam Shukla is a Professor in the ICT Department of the institute. He has 19 years of teaching experience. His research interest includes Speech processing, Artificial Intelligence, Soft Computing and Bio-informatics. Prior to IIITM, Gwalior, he was Reader and Head at NIT, Raipur in Biomedical Engineering Department.

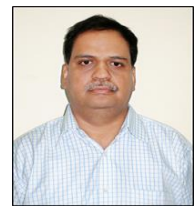

Gaurav Ojha is a student in the Department of Information Technology at Indian Institute of Information Technology and Management, Gwalior. His areas of interest are Web Technologies, Open Source Software and Internet Security.

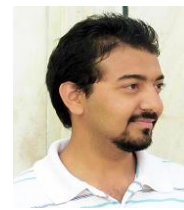

Sachin Acharya is a student in the Department of Information Technology and Indian Institute of Information Technology and Management, Gwalior. His areas of interest are OS, Optimization methods and algorithms.

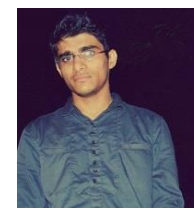

Shubham Jain isa student in the Department of Information Technology, Indian Institute of Information Technology and Management, Gwalior. He is interested in a variety of nature inspired algorithms and optimization techniques.

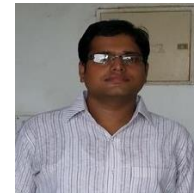